\documentclass[%
 reprint,
superscriptaddress,
nofootinbib,
 amsmath,amssymb,
 aps,
prd,
]{revtex4-2}

\usepackage{xcolor}
\usepackage{graphicx}% Include figure files
\usepackage{dcolumn}% Align table columns on decimal point
\usepackage{bm}% bold math
\usepackage[hidelinks]{hyperref}% add hypertext capabilities
\usepackage[mathlines]{lineno}% Enable numbering of text and display math
\usepackage[normalem]{ulem}
\usepackage{marginnote}
\usepackage[commentmarkup=uwave]{changes}
\definechangesauthor[name={Aaron Johnson}]{AJ}

\usepackage{tikz}

\definecolor{mpl_blue}{HTML}{1F77B4}
\definecolor{mpl_orange}{HTML}{FF7F0E}
\definecolor{mpl_green}{HTML}{2CA02C}
\definecolor{mpl_red}{HTML}{D62728}

\newcommand{\CIT}{\affiliation{TAPIR, California Institute of Technology, Pasadena, CA 91125, USA}}
\newcommand{\CITLab}{\affiliation{LIGO Laboratory, California Institute of Technology, Pasadena, California 91125, USA}}

\begin{document}
\include{macros}

%\preprint{APS/123-QED}

% \title{\textsc{PETRA}: From the global fit for LISA's Galactic binaries to a catalog of sources}
\title{\textsc{PETRA}: From the LISA global fit to a catalog of Galactic binaries}

\author{Aaron~D.\ Johnson} 
\email{Corresponding author: aaronj@caltech.edu} 
\CIT

\author{Javier Roulet} 
%\email{jroulet@caltech.edu} 
\CIT

\author{Katerina Chatziioannou} 
%\email{kchatziioannou@caltech.edu} 
\CIT 
\CITLab

\author{Michele Vallisneri} 
%\email{vallis@caltech.edu} 
\affiliation{ETH Zurich, Institute for Particle Physics and Astrophysics,
Wolfgang-Pauli-Str.~27, 8093 Zurich, Switzerland}
% \affiliation{Jet Propulsion Laboratory, California Institute of Technology, Pasadena CA 91109, USA}

\author{Chris G.\ Trejo} 
%\email{chris@trejocosmico.com} 
\CITLab

\author{Kyle A. Gersbach} 
%\email{kyle.a.gersbach@vanderbilt.edu} 
\affiliation{Department of Physics \& Astronomy, Vanderbilt University, 2301 Vanderbilt Place, Nashville, TN 37235, USA}

\begin{abstract}
The Laser Interferometer Space Antenna (LISA) will detect mHz gravitational waves from many astrophysical sources, including millions of compact binaries in the Galaxy, thousands of which may be individually resolvable.
The large number of signals overlapping in the LISA dataset requires a \emph{global fit} in which an unknown number of sources are modeled simultaneously.
This introduces a \emph{label-switching ambiguity} for sources in the same class, making it challenging to distill a traditional astronomical catalog from global-fit posteriors.
We present a method to construct a catalog by post-processing the global-fit posterior,
% optimally
relabeling samples to minimize the statistical divergence between the global fit and a factorized catalog representation.
The resulting catalog consists of the source posterior distributions and their probabilities of having an astrophysical origin.
We demonstrate our algorithm on two toy models and on a small simulated LISA dataset of Galactic binaries.
Our method is implemented in the open-source Python package \textsc{petra\_catalogs}, and it can be applied in postprocessing to the output of any global-fit sampler.
\end{abstract}

\maketitle

%%%%%%%%%%%%%%%%%%%%%%%%%%%%%%%%%
\section{Introduction}
%%%%%%%%%%%%%%%%%%%%%%%%%%%%%%%%%

The first decade of gravitational-wave astronomy (2015--2025) has yielded dozens of detections of stellar-mass binary mergers from ground-based detectors operating around 100\,Hz~\cite{LIGOScientific:2018mvr, LIGOScientific:2020ibl, KAGRA:2021vkt,Nitz2023,Wadekar2023}, as well as evidence for a nHz stochastic background (likely from supermassive black-hole binaries) from pulsar timing arrays~\cite{NANOGrav:2023gor, reardon_search_2023, epta_collaboration_and_inpta_collaboration_second_2023}.
The Laser Interferometer Space Antenna (LISA), planned for launch around 2035, will detect mHz gravitational waves from a variety of sources, including massive black-hole binary mergers, extreme-mass ratio inspirals, stellar-origin black-hole binaries, and compact binaries in the Galaxy (mostly double white-dwarf binaries) \cite{Colpi:2024xhw}.
Galactic binaries will be by far the most plentiful source that LISA will observe: tens of millions will crowd the LISA band, with thousands individually resolvable \cite{Littenberg:2023xpl} and the rest adding incoherently to create a confusion foreground, \cite{hils_gravitational_1990, bender_confusion_1997}.

Because so many gravitational-wave sources will be superimposed in the LISA data, its analysis is envisioned as a \emph{global fit}, in which all astrophysical sources (and detector noise) are modeled simultaneously~\cite{Cornish:2005qw,Robson:2017ayy,Littenberg:2023xpl,Katz:2024oqg,Strub:2024kbe}.
The number $N$ of resolvable sources will be unknown \textit{a priori}, requiring methods to assess models with different numbers of sources, such as transdimensional inference approaches \cite{Umstatter:2005su}.
Furthermore, any statistical model of the dataset will be subject to a \emph{label-switching ambiguity} with respect to sources of the same type. Consider for instance Galactic binaries, and let binary $\alpha$ have gravitational waveform $h(\theta_\alpha)$, with $\theta_\alpha$ the binary parameters. Schematically, the LISA data will be $d = n + \sum_\alpha h(\theta_\alpha)$, where $n$ is instrument noise; then the posterior has the form~\cite{greenReversibleJumpMarkov1995}
\begin{multline}
\label{eq:globalfit}
p_\mathrm{gf}(N; \{\theta_1, \ldots, \theta_N\} \mid d) \\
= p_\mathrm{gf}(N \mid d) \, p_\mathrm{gf}(\{\theta_1, \ldots, \theta_N \} \mid N, d)\,,
\end{multline}
where the parameters in braces comprise an \emph{unordered set}, since any permutation of the indices would yield the same total signal.

The label-switching ambiguity is an instance of model non-identifiability and is a commonly occurring problem in statistical inference, for example with mixture models \cite{Stephens:2000a, Jasra:2005a}.
Proposed solutions include applying identifiability constraints (i.e., \textit{a priori} assumptions on the parameters such as parameter orderings) to resolve the ambiguity, using clustering algorithms on posterior samples, adopting decision-theoretic approaches that augment the statistical model with a labeling loss function, and others yet \cite{Stephens:2000a,Jasra:2005a,Geweke:2007a,Puolamaki:2009a}.
For ground-based detectors, a label-switching ambiguity writ smaller exists for the two components of a binary.
It is typically addressed by ordering the components by their masses, but spin sorting \cite{Biscoveanu:2020are}, posterior clustering \cite{Gerosa:2024ojv}, and tidal sorting \cite{Golomb:2024mmt} have also been proposed.

In the LISA context, the label-switching ambiguity means that a global fit does not immediately define a traditional astronomical \emph{catalog}.
In our definition, a catalog of $M$ sources consists, for each source individually, of a probability density $p_\alpha(\theta)$ for the source's physical parameters and of the source's probability $P^\ast_\alpha$ of being real under the astrophysical model underlying the global fit.
If sources were distinguishable, then catalog entries would be defined by the marginal posteriors $p_\mathrm{gf}(\theta_\alpha \mid d)$. 
Because they are indistinguishable, all such marginals are statistically identical, and they are useless (say) for electromagnetic source identification or follow up \cite{Littenberg:2019mob,Littenberg:2024bso}.\footnote{A catalog $\{p_\alpha(\theta),P^\ast_\alpha\}$ would still admit a \emph{global} label-switching invariance, which can be resolved trivially by parameter ordering.}

In this paper we propose a method to construct a catalog from a global-fit posterior represented by a chain of (potentially transdimensional) Markov Chain Monte Carlo samples.
Our method is inspired by Stephens' algorithm for Gaussian-mixture identification \cite{Stephens:2000a}.
Loosely speaking, we seek to obtain the most ``catalog-like'' relabeling of $p_\mathrm{gf}$: that is, the assignment of each unordered parameter set to labeled catalog sources that minimizes the statistical ``distance'' between $p_\mathrm{gf}$ and the factorized catalog distribution.
The method is based on two operations:
\begin{enumerate}
    \item The catalog source distributions $p_\alpha(\theta)$ are the marginal distributions of the relabeled global fit, and the probability $P^\ast_\alpha$ of a source being real is the fraction of posterior samples that contain it.
    This is similar, but not fully equivalent, to the evidence for global-fit models with different numbers of sources, which is proportional to the fraction of global-fit samples for which that model is active~\cite{Lodewyckx:2011a,littenbergBayesianApproachDetection2009a,Littenberg:2020bxy}.
    \item Given $p_\alpha(\theta)$ and $P^\ast_\alpha$, the entries of each posterior sample are ``sorted'' among sources such that the probability that they correspond to that source is maximized.
\end{enumerate}

We converge to the optimal relabeling iteratively: we define a $\theta$-dependent relabeling $\ell$ that maps $p_\mathrm{gf}$ to a $p'_\mathrm{rel}$ that is \emph{not} invariant to label switching, and then fit $p'_\mathrm{rel}$ with an auxiliary parametric distribution $q'_\phi$ that factorizes over sources (the primes denote distributions of ordered parameter vectors with the possibility of null entries, as explained in Sec.~\ref{sec:formalism}).
We then alternately optimize the relabeling $\ell$ and the fit parameters $\phi$ to maximize the similarity of $p'_\mathrm{rel}$ and $q'_\phi$, as measured by the Kullback--Leibler divergence~\cite{Kullback:1951zyt}.
The final catalog entries $\{p_\alpha(\theta),P^\ast_\alpha\}$ are obtained from the single-source marginal posteriors of $p'_\mathrm{rel}$.

Since we have obtained this catalog from the global fit samples, we do not annotate $\{p_\alpha(\theta),P^\ast_\alpha\}$ as conditioned directly on the data.
Moreover, the global fit cannot be reconstructed from the catalog, since the information about the correlation between sources is lost.
In the event that correlations are important, one may produce a catalog that includes composite ``sources,'' which lack individual identity and contain a multiple of an individual source's number of parameters, but we do not consider such a catalog definition here.

Our method, \textsc{Petra}, is implemented in the open-source Python package \textsc{petra\_catalogs}~\cite{johnson_petra_2025}, and it is presented in detail in Sec.~\ref{sec:methods}.
In Sec.~\ref{sec:toy} we apply the method to two toy problems, involving reshuffled normal distributions and superimposed sinusoids.
In Sec.~\ref{sec:globalfit} we apply the method to a LISA global-fit posterior for a small number of Galactic binary sources.
In Sec.~\ref{sec:conclusions} we present our conclusions and compare our method to previously proposed solutions for the LISA label-switching ambiguity, including hierarchical Ward clustering \cite{Umstatter:2005su}, source-ordering parameter transformations \cite{Buscicchio:2019rir}, and heuristic clustering based on waveform similarity \cite{Littenberg:2020bxy,Lackeos:2023eub}.

%%%%%%%%%%%%%%%%%%%%%%%%%%%%%%%%%%%%%%%%%
\section{From global fit to catalog}
\label{sec:methods}
%%%%%%%%%%%%%%%%%%%%%%%%%%%%%%%%%%%%%%%%

In this section we describe how we construct a catalog from a global fit.
In Sec.~\ref{sec:formalism} we introduce our formalism and motivate the method; in Sec.~\ref{sec:algorithm} we present a practical algorithmic implementation; and in Sec.~\ref{sec:information-theory} we provide an alternative information-theoretical motivation.

%%%%%%%%%%%%%%%%%%%%%%%%%%%%%%%
\subsection{Formalism}
\label{sec:formalism}

A \emph{global fit} for a class of signals $h(\theta)$ is the joint posterior density over the number of sources $N$ and the source parameters $\Theta_N$, Eq.~\eqref{eq:globalfit}.
Since the global fit is invariant to reordering the source parameters, we represent them as unordered sets (point clouds)
\begin{equation}
    \Theta_N = \{\theta_1, \ldots, \theta_N\} \in \Omega^{(N)}\,,
\end{equation}
where $\Omega$ is the space of physical parameters of a single source and $\Omega^{(N)}$ is the space of $N$-point clouds (unordered sets of $N$ elements of $\Omega$).
The global fit $p_\mathrm{gf}(\Theta)$ is a probability distribution over $\bigcup_N\Omega^{(N)}$.\footnote{In practice most stochastic samplers use ordered tuples to represent the phase space, producing $N!$ equivalent solutions that result from permuting the indices.}
For brevity, we will use
\begin{equation}
    \Theta = (N, \Theta_N)\,,
\end{equation} 
and drop the conditioning on the data---for example, the global fit is $p_\textrm{gf}(\Theta)$.

From the global fit, we construct a catalog of $M$ sources, with $M \geq N$, at least as large as the maximum number of entries in the global fit.\footnote{This is a requirement of the formalism; a smaller catalog may be produced \textit{a posteriori} by excluding sources with a low probability of being astrophysical.}
We introduce a labeling rule $\ell$ that assigns the entries of $\Theta$ among the $M$ distinct sources that comprise the catalog.
For any $\Theta_N$, $\ell$ produces an ordered tuple of size $M$,
\begin{equation}
    \ell(\Theta) = \bm \theta' \equiv (\theta'_1, \ldots, \theta'_M)\,,
\end{equation}
of which $N$ elements are physical sources ($\theta'_\alpha \in \Omega$) and the remaining $M-N$ take a null value (which we denote as $\theta'_\alpha = \varnothing$).
Formally,
\begin{align}
    \ell &: \bigcup_{N=0}^M \Omega^{(N)} \to \Omega'^M\,, \\
    \Omega' &= \Omega \cup \{\varnothing\}\,,
\end{align}
where $\Omega'$ is the union of the physical parameter space $\Omega$ and the null value, and $\Omega'^M$ is the space of $M$-tuples (ordered sequences of $M$ elements of $\Omega'$).
Throughout this paper we use primes for hybrid continuous--discrete objects (whether variables or probabilities) that can represent source parameter sets or null values.
In summary, $\Theta \in \bigcup_N\Omega^{(N)}$ represents a label-switching-symmetric object (a global-fit sample), while $\bm\theta' \in \Omega'^M$ represents an ordered object (a relabeled sample) obtained from $\Theta$ via an invertible labeling rule $\ell$.

The distribution of relabeled sources $p'_\mathrm{rel}(\bm\theta')$ is obtained from the change of variables formula as
\begin{equation}
    \label{eq:p_rel}
    p'_\mathrm{rel}(\bm \theta') = \begin{cases}
        p_\mathrm{gf}(\ell^{-1}(\bm\theta')) & \text{if $\exists\,\Theta : \ell(\Theta) = \bm \theta'$}\,,\\
        0 & \text{if not}\,,
    \end{cases}    
\end{equation}
which follows from the fact that $\ell$ is an invertible labeling operation with unit Jacobian determinant.
This density is called a pushforward $p'_\mathrm{rel} = \ell_\ast p_\mathrm{gf}$ of the global fit by the labeling rule \cite{Kobyzev:2021a}.
%\textcolor{red}{\marginpar{(11)}(because each entry can be assigned to only one source)}

We construct the catalog as the product of the relabeled distribution marginals:
\begin{equation}
    \label{eq:p_catalog_general}
    p'_\mathrm{cat}(\bm \theta') = \prod_{\alpha=1}^M p'_\alpha(\theta'_\alpha)\,,
\end{equation}
where the marginals are
\begin{align}
    p'_\alpha(\theta'_\alpha)
    &\equiv \int \mathrm{d}\bm\theta'_{\setminus\alpha} \, p'_\mathrm{rel}(\bm\theta') \\
    &= \begin{cases}
         P^\ast_\alpha \, p_\alpha(\theta'_\alpha) & \text{if $\theta'_\alpha \in \Omega$}\,,\\
        1 - P^\ast_\alpha & \text{if $\theta'_\alpha = \varnothing$}\,.
    \end{cases}
    \label{eq:catdefine}
\end{align}
In the first line, the integration is performed for all indices other than $\alpha$. 
For each source, Eq.\ \eqref{eq:catdefine} defines the \emph{probability of astrophysical origin}
\begin{equation}
    P^\ast_\alpha
        \equiv P_\alpha(\theta'_\alpha {\in} \Omega)
        = 1 - P_\alpha(\theta'_\alpha {=} \varnothing)\,,
\end{equation}
and the \emph{catalog posterior}
\begin{equation}
    p_\alpha(\theta'_\alpha) \equiv p_\alpha(\theta'_\alpha \mid \theta'_\alpha{\in}\Omega)\,.
\end{equation}
The probability of astrophysical origin is conditional on the astrophysical model underlying the construction of the global fit and on the labeling assigned by our algorithm.
The catalog consists of $M$ pairs of parameter posteriors and probabilities of astrophysical origin $\big\{\big(p_\alpha(\theta), P^\ast_\alpha\big)\big\}_{\alpha=1}^M$. The joint catalog probability is then
\begin{equation}
\label{eq:p_catalog}
    p'_\mathrm{cat}(\bm \theta')
    = \bigg[
    \prod_{\alpha:\theta'_\alpha \in \Omega}P^\ast_\alpha \, p_\alpha(\theta'_\alpha)
    \bigg]
    \bigg[
    \prod_{\beta:\theta'_\beta = \varnothing} \big(1 - P^\ast_\beta\big)
    \bigg]\,.
\end{equation}

In summary, the catalog is obtained from the global fit using the labeling rule $\ell$.
For the purpose of this work, the global fit is an input and we seek a labeling rule that maximizes the assignment probability $p'_\mathrm{cat}(\bm \theta')$, as we explain further in Sec.~\ref{sec:algorithm}.
In Sec.~\ref{sec:information-theory} we demonstrate that this optimization criterion also minimizes the information loss between $p'_\mathrm{cat}$ and $p'_\mathrm{rel}$, as well as their Kullback--Leibler divergence.

%%%%%%%%%%%%%%%%%%%%%%%%%%%%%%%
\subsection{Algorithm}
\label{sec:algorithm}

Stochastic samplers represent the global fit as a chain of $S$ samples $\{\Theta^i\}_{i = 1}^S$.
For each sample, we seek the labeling $\ell^i$ that maximizes the probability that the ordered sample was drawn from $p'_\mathrm{cat}$ (assignment probability), which amounts to maximizing $p'_\mathrm{cat}(\ell(\Theta^i))$.
The catalog $p'_\mathrm{cat}$ naturally incorporates a ``penalty" of $P^\ast$ for assigning an entry to a source with $P^\ast < 1$, and of $1-P^\ast$ for leaving a source with $P^\ast > 0$ unassigned.

The computation of $p'_\mathrm{cat}$ via Eq.~\eqref{eq:p_catalog} requires knowledge of the catalog components $\{(p_\alpha(\theta), P^\ast_\alpha)\}_{\alpha=1}^M$, which depend on the labeling rule itself and moreover have no analytical form.
We estimate these quantities iteratively, starting from an initial labeling rule (say, random assignment).
For each sample, we choose the  labeling for the next iteration by maximizing
\begin{equation}
\label{eq:p_catalog_approx}
    q'_\phi(\bm \theta')
    = \bigg[
        \prod_{\alpha:\theta'_\alpha \in \Omega} Q^\ast_\alpha q(\theta'_\alpha \mid \phi_\alpha)
    \bigg]
    \bigg[
        \prod_{\beta:\theta'_\beta = \varnothing} \big(1 - Q^\ast_\beta\big)
    \bigg]\,,
\end{equation}
where we have replaced $p'_\mathrm{cat}$ (which is after all the \emph{endpoint} of this algorithm) with an approximation in terms of $Q^\ast_\alpha$ (a Monte Carlo estimate of $P^\ast_\alpha$ as the fraction of samples that have that source assigned) and of a parametric ansatz $q(\theta_\alpha \mid \phi_\alpha)$ for the source posteriors $p_\alpha(\theta_\alpha)$.
In this paper we adopt the simple multivariate normal
\begin{equation}
    q(\theta_\alpha \mid \phi_\alpha) = \mathcal N(\theta_\alpha \mid \mu_\alpha, \Sigma_\alpha)\,.
    \label{eq:gaussian}
\end{equation}

We maximize Eq.~\eqref{eq:p_catalog_approx} iteratively, alternating between estimating the probabilities of astrophysical origin $Q^\ast_\alpha$ and fit parameters $\phi_\alpha$ based on the current labeled entries, and selecting the optimal relabelings $\ell(\Theta^i)$ based on the current estimates of $Q^\ast_\alpha$ and $\phi_\alpha$.
Keeping the labeling fixed, we maximize a reward function $R$, defined as the logarithm of the joint probability of drawing all the samples assigned to each source $\alpha$ from $q'_\phi$ [Eq.~\eqref{eq:p_catalog_approx}]:
\begin{equation}
\label{eq:log_joint_prob}
    R\big(\{Q_\alpha^\ast, \mu_\alpha, \Sigma_\alpha\}_{\alpha=1}^M\big)
    = \sum_{i=1}^S \log q'_\phi(\bm\theta'^i)\,.
\end{equation}
Equation~\eqref{eq:log_joint_prob} is maximized analytically by setting its derivatives with respect to the optimization parameters $\{Q_\alpha^\ast, \mu_\alpha, \Sigma_\alpha\}$ to zero.
Then the optimal $Q^\ast_\alpha$ are obtained straightforwardly as the fraction of samples that contain a source, whereas fitting the $q$'s amounts to computing the means and covariances (over samples) of each $\theta'^i_\alpha = (\ell(\Theta^{i}))_\alpha$ for the current relabeling.
The relabeling of each sample is performed by selecting one value from each row and column of the reward matrix $\mathbf{R}^i \in \mathbb{R}^{M\times M}$, constructed as
\begin{equation}
    R^i_{\alpha\beta} 
    =
    \begin{cases}
        \log \big[Q^\ast_\alpha \, q(\theta'^i_\beta \mid \phi_\alpha)\big] & \text{if $\theta'^i_\beta \in \Omega$}\,, \\
        \log(1 - Q^\ast_\alpha) & \text{if $\theta'^i_\beta = \varnothing$}\,,
    \end{cases}
    \label{eq:reward}
\end{equation}
such that their sum is maximized.\footnote{There is a technical issue with this procedure: if at some iteration a source is fully assigned ($Q^\ast_\alpha = 1$) or unassigned ($Q^\ast_\alpha=0$), there is an infinitely large penalty to any proposed labeling that changes that status. We mitigate this by clipping $Q^\ast$ between $10^{-6}$ and $1-10^{-6}$. We have verified that a more aggressive clipping of $10^{-2}$ yields similar results.}
This step amounts to assigning each global-fit entry to a catalog source for each stochastic sample.
The \emph{Hungarian algorithm} finds the optimal relabeling in $\mathcal{O}(M^3)$ operations \cite{Kuhn:1955a}; in practice we use the faster Jonker--Volgenant algorithm \cite{Crouse:2016a} implemented in \textsc{SciPy} as \textsc{linear\_sum\_assignment} \cite{Virtanen:2019joe}.
Both steps of the algorithm are trivially parallelizable: fitting the parametric ansatz can be parallelized over sources, while labeling the sources can be parallelized over samples.

\begin{figure*}
    \centering
    \includegraphics[width=\textwidth]{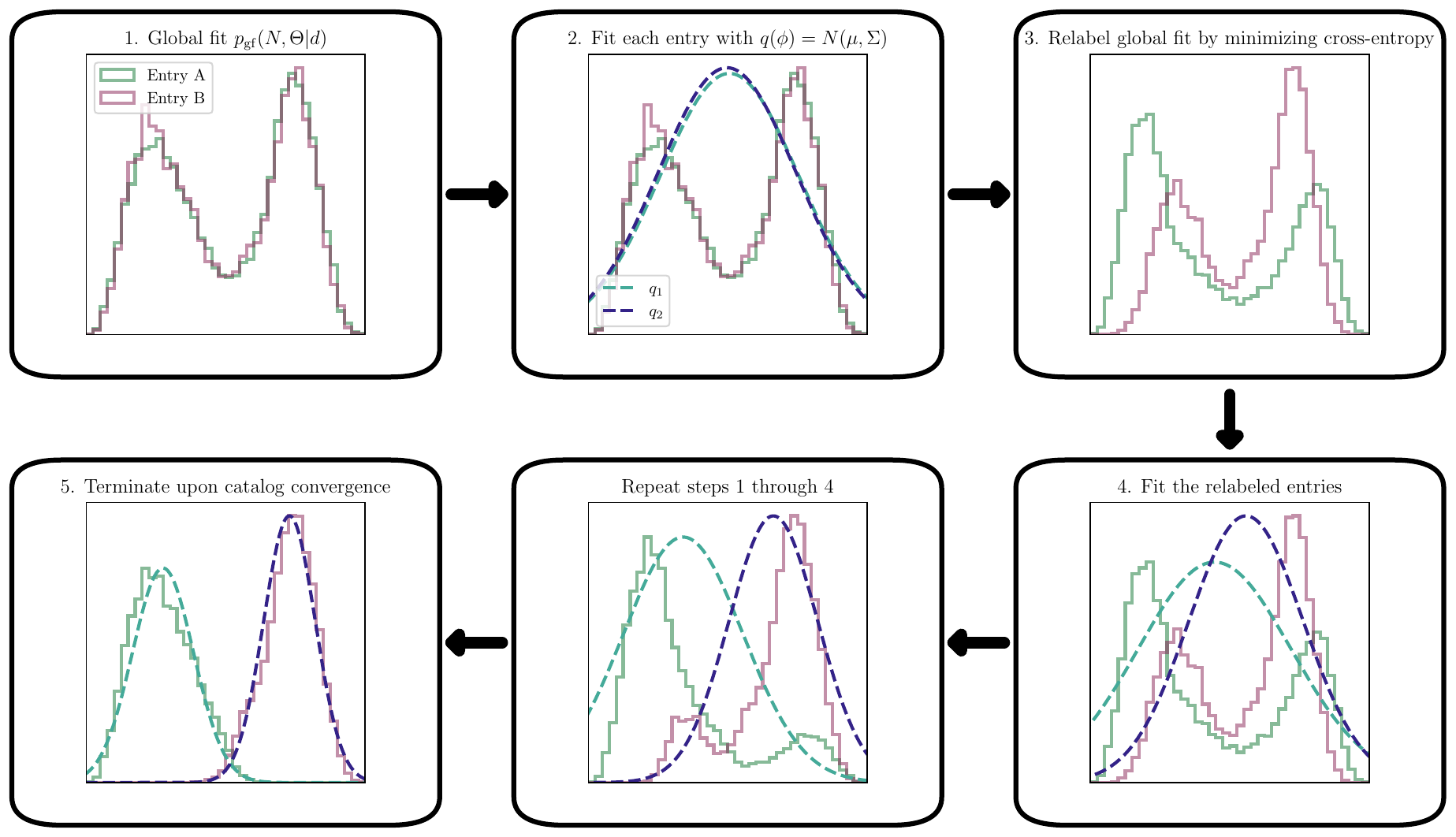}
    \caption{Illustration of the global-fit to catalog algorithm, for a toy model of two sinusoidal signals parameterized by amplitude, frequency, and phase (see Sec.~\ref{sec:toy-sinusoid}). For clarity we show only marginal posteriors of frequency, although the algorithm makes use of all parameters.
    Panel 1: We begin by identifying the relabeled posterior with the global fit. For fully converged chains all the marginal posteriors are statistically identical.
    Panel 2: We fit the marginals of the relabeled posterior with a multivariate normal over all source parameters.
    Panel 3: For every sample, we reassign each entry to either source (i.e., either normal) by maximizing the assignment probability, Eq.\ \eqref{eq:p_catalog_approx}.
    Panel 4: We repeat steps 2 and 3 until the normals and relabelings stop changing. The marginals of the final relabeled samples define the two catalog posteriors.
    \label{fig:workflow}}
\end{figure*}

The algorithm terminates when the cost function remains unchanged after the ansatz re-fit and relabeling.
Since at each step the reward function $\sum_i \log q'_\phi(\bm \theta'^i)$ is maximized (either with respect to its parameters or the relabeling), it can never decrease from its previous value. There is a finite number of ways of relabeling the set of global-fit samples, so the algorithm always converges to a (local) maximum of the reward function.
The catalog posteriors $p_{\alpha}$ (i.e., the marginals of the relabeled global fit) are represented directly by ``columns'' of the stochastic-sampler chain under the final relabeling.
We illustrate the process in Fig.~\ref{fig:workflow} for the simple case of two sinusoidal signals that result in uncorrelated quasinormal posteriors.

The process is deterministic given a starting list of global fit samples.
To reduce dependence on how the global fit samples are initially ordered, we start by randomly shuffling the global fit entries, including the null values, for each global fit sample. 
Convergence is aided by a heuristic initialization step.
For fixed-dimensionality global fits, we histogram a single parameter across all samples and entries, find the mode, and attribute the closest entry in each sample to (interim) source 1. Then we create a new histogram from all unattributed entries, and repeat.
For transdimensional global fits, we carry out the full iterative procedure with a univariate normal ansatz and use the result to initialize the full multivariate normal subsequent calculation.

This method, which we named \textsc{Petra}\footnote{Posterior Estimate Transformation Resolving Ambiguity, but really KC's cat.}, is implemented in the Python package \textsc{petra\_catalogs}~\cite{johnson_petra_2025}.
\textsc{Petra} can load posterior chains in the formats output by \textsc{PTMCMCSampler} \cite{Ellis:2017b} and \textsc{UCBMCMC} \cite{ldasoft}, as well as a generic tabular format, for both fixed and variable numbers of sources.
Using the initialization step described above, \textsc{Petra} can relabel 50,000 samples from a 10-source, 8-parameter model in less than 2 minutes.
The code has a modular structure that can support other parametric ansatz distributions and cost functions.
Relabeled posteriors are output in the same format as the input posterior chains.

%%%%%%%%%%%%%%%%%%%%%%%%%%%%%%%%%%%
\subsection{Information-theoretic interpretation}
\label{sec:information-theory}

In Sec.~\ref{sec:algorithm} we motivated this procedure as finding the labeling that maximizes the ``assignment'' probability that the entries in each sample were drawn from specific sources in the catalog.
The same objective also results from minimizing the information lost by discarding the correlations between sources when building the catalog as a product of marginals, Eq.~\eqref{eq:p_catalog_general}.
The information loss is equal to the difference of the entropies of $p'_\mathrm{rel}$ and $p'_\mathrm{cat}$:
\begin{align}
    I_{\rm loss}
    = - H(p'_\mathrm{rel}) + \sum_{\alpha=1}^M H(p'_\alpha) \geq 0\,.
    \label{eq:info_loss}
\end{align}
For the case of two sources, Eq.~\eqref{eq:info_loss} would be called the mutual information $I(\theta'_1; \theta'_2)$.
We seek the relabeling $\ell$ that minimizes $I_{\rm loss}$.

By Eq.~\eqref{eq:p_rel}, the entropy of $p'_\mathrm{rel}$ equals that of $p_\mathrm{gf}$, independently of the relabeling rule:
\begin{align}
    H(p'_\mathrm{rel})
    &\equiv -\int p'_\mathrm{rel}(\bm\theta') \log p'_\mathrm{rel}(\bm\theta') \, {\rm d}\bm\theta' \label{eq:h_p_rel}\\
    % &= -\int p_\mathrm{gf}(\Theta) \log p'_\mathrm{rel}(\ell(\Theta)) \, {\rm d}\Theta \\
    &= -\int p_\mathrm{gf}(\Theta) \log (p_{\rm gf}(\Theta)) \, {\rm d}\Theta \\
    &\equiv H(p_\mathrm{gf})\,,
\end{align}
where the integrals over some function $f$ are meant as
\begin{align}
    \int\mathrm{d}\theta' f(\theta') &\equiv f(\varnothing) + \int_\Omega \mathrm d\theta \, f(\theta)\,, \\
    \int \mathrm{d}\Theta \, f(\Theta) &\equiv \sum_{N=0}^M \int_{\Omega^{(N)}} \mathrm{d}\Theta_N \, f(N, \Theta_N)\,.
\end{align}
Since $H(p'_\mathrm{rel})$ is independent of the relabeling rule, minimizing $I_{\rm loss}$ amounts to minimizing
\begin{align}
    \sum_\alpha H(p'_\alpha)
    &= - \sum_\alpha\int p'_\alpha(\theta'_\alpha) \log p'_\alpha(\theta'_\alpha) \, {\rm d}\theta'_\alpha \\
    &= - \sum_\alpha\int p'_\mathrm{rel}(\bm\theta') \log p'_\alpha(\theta'_\alpha) \, {\rm d}\bm\theta' \\
    &= - \int p'_\mathrm{rel}(\bm\theta') \log p'_\mathrm{cat}(\bm\theta') \, {\rm d}\bm\theta' \label{eq:cross_entropy}\\
    &= - \int p_{\rm gf}(\Theta) \log p'_\mathrm{cat}(\ell(\Theta)) \, {\rm d}\Theta \\
    &\approx -\frac{1}{S} \sum_{i=1}^S \log p'_\mathrm{cat}(\ell(\Theta^i)); \quad \Theta^i \sim p_\mathrm{gf}\,.
    \label{eq:loss_function}
\end{align}
\textsc{Petra} minimizes the loss function of Eq.~\eqref{eq:loss_function}, which (by Eq.~\eqref{eq:cross_entropy}) equals the cross-entropy between the relabeled global fit $p'_\mathrm{rel}$ and the catalog $p'_\mathrm{cat}$.
Indeed, $I_\mathrm{loss}$ is just the Kullback--Leibler divergence between the two distributions.
In this sense, we are selecting the relabeling that makes $p'_\mathrm{rel}$ closest to the product of its marginals $p'_\mathrm{cat}$.

%%%%%%%%%%%%%%%%%%%%%%%%%%%%%%%%%%%%%
\section{Toy models}
\label{sec:toy}
%%%%%%%%%%%%%%%%%%%%%%%%%%%%%%%%%%%%

In this section we demonstrate \textsc{Petra} on two toy examples. 
In Sec.~\ref{sec:toy-gaussian} we start with two Gaussian bidimensional distributions, shuffle posterior samples between them, and show that \textsc{Petra} relabels the samples accurately even when the original distributions are very similar. 
In Sec.~\ref{sec:toy-sinusoid} we consider data consisting of a sum of sinusoidal signals, including two sinusoids with increasing frequency offset; two sinusoids, where one has a bimodal posterior; and ten sinusoids, with only five detectable.
In all cases \textsc{Petra} creates catalogs with posteriors that agree with the true parameters of detectable sources, and with appropriate probabilities of astrophysical origin.

%%%%%%%%%%%%%%%%%%%%%%%%%%%%%%%%%%%%%
\subsection{Overlapping Gaussian posteriors}
\label{sec:toy-gaussian}
%%%%%%%%%%%%%%%%%%%%%%%%%%%%%%%%%%%%%

\begin{figure*}
    \centering
    \includegraphics[width=0.99\textwidth]{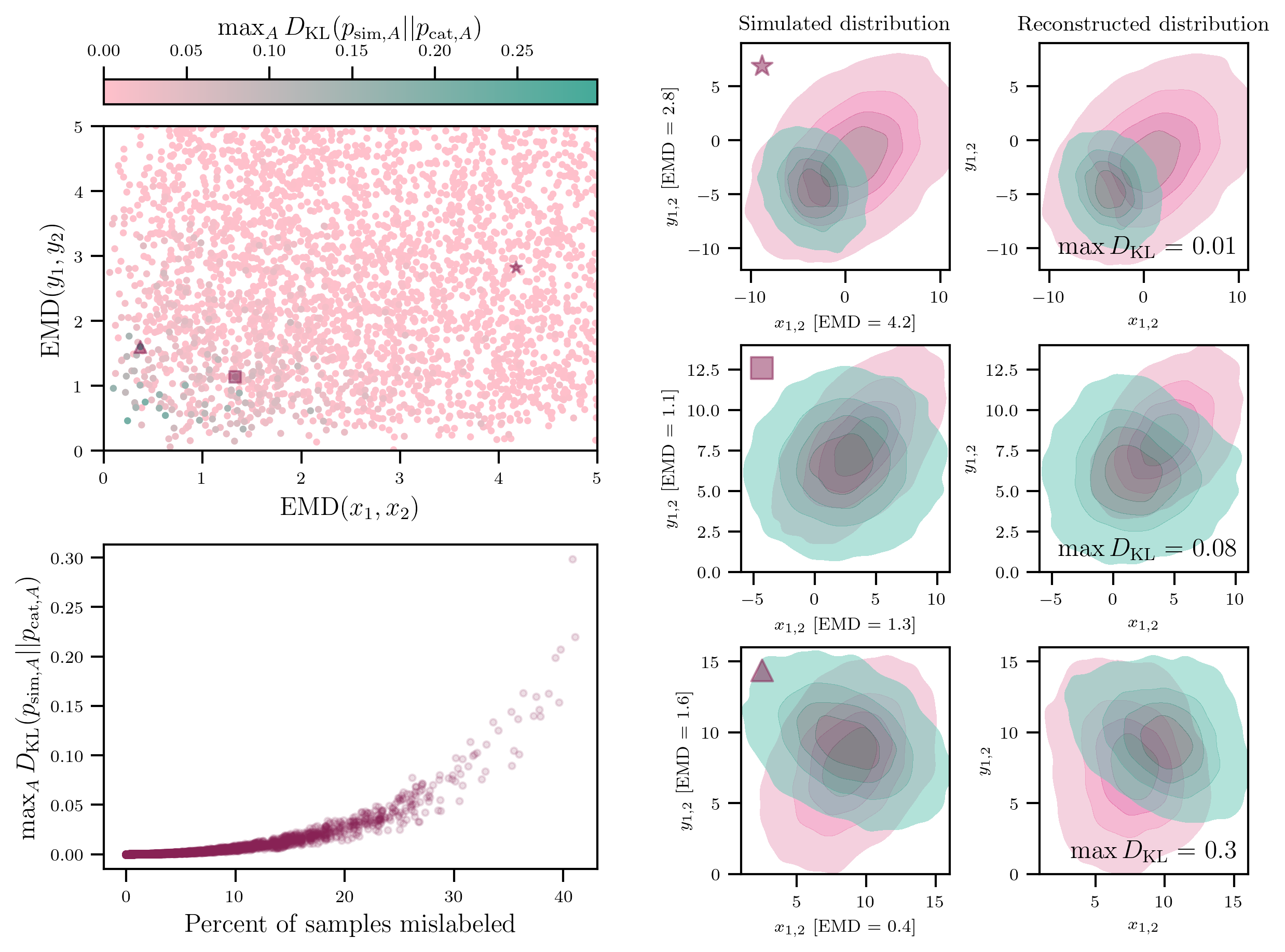}
    \caption{\textsc{Petra} relabeling of two overlapping bidimensional Gaussians (first test of Sec.\ \ref{sec:toy-gaussian}).
    \emph{Top left}: KL divergence ($\max_A D_\mathrm{KL}(p_{\mathrm{sim},A}||p_{\mathrm{cat},A})$, color scale) between simulated and reconstructed distributions, plotted against 1D EMDs of the variable pairs $x_{1,2}$ and $y_{1,2}$. Smaller EMDs correspond to greater overlaps: distributions begin to overlap at EMD $\sim 5$, and at EMD $\sim 2.5$ the mean of one falls within the support of the other.
    \emph{Bottom left}: KL divergence versus fraction of mislabeled samples. A large fraction of mislabelings in a region of strong overlap will not affect the divergence significantly, since the samples are effectively indistinguishable. 
    \emph{Top, middle, and bottom left}: Examples of simulated and reconstructed distributions corresponding to the star, square, and triangle in the top left plot. Reconstructed and injected distributions appear very close even for low EMDs.
    \label{fig:fake_posterior_kl_div}}
\end{figure*}

Our first test involves 16,000 pairs of bidimensional Gaussian distributions $p_{\mathrm{sim},A}(x_A, y_A) = \mathcal{N}(\boldsymbol{\mu}_A, \mathbf{\Sigma}_A)$, for $A=\{1, 2\}$.
The means and covariances are drawn randomly as $\boldsymbol{\mu}_{A,i} \sim U(-10, 10)$ and $\Sigma_{A,ij} = (C_{A,ij} + C_{A,ji}) / 2$, with 
$C_{A,ij} \sim \mathcal{N}(0, \sigma_A)$ and $\sigma_A \sim U(1, 10)$.
Non-positive-definite covariance matrices are rejected.
For each pair of distributions, we draw 10,000 random samples and shuffle the pairs to create a joint label-switching-invariant ``global fit.'' 
We then apply the algorithm of Sec.~\ref{sec:methods} to build the two catalog distributions, and compare them to the original unshuffled distributions. 
We quantify their difference as $\max_A D_\mathrm{KL}(p_{\mathrm{sim},A}||p_{\mathrm{cat},A})$, where the Kullback--Leibler divergence $D_\mathrm{KL}$ is computed following \cite[Eq.~(3)]{Zhang:2021a}.
We characterize the difficulty of each relabeling problem by Wasserstein ``earth mover'' distances (EMD, \cite{Rubner:1998a}) between the two pairs of simulated marginalized distributions $p_\mathrm{sim}(x_1)$ and $p_\mathrm{sim}(x_2)$, and $p_\mathrm{sim}(y_1)$ and $p_\mathrm{sim}(y_2)$. We compute these one-dimensional EMDs using \textsc{SciPy} \cite{Virtanen:2019joe}.

Results are collected in Fig.~\ref{fig:fake_posterior_kl_div}. The top left panel plots the divergence as the color scale on the plane defined by the two EMDs.
The $p_\mathrm{sim}$ begin to overlap along each axis for EMD $\sim 5$. 
As the overlaps increase ($\mathrm{EMD} \rightarrow 0$), so does the divergence, indicating that the $p_\mathrm{sim}$ are recovered less accurately. 
The bottom left panel shows that larger numbers of catalog samples are mislabeled for larger divergence.
This is not a shortcoming of the algorithm, but the expected outcome in the confusion limit.

The right panel compares simulated and reconstructed distributions for three examples, identified by the star, square, and triangle in the left panel.
In the top panel (star), the simulated distributions overlap fully but have different variances, with EMD$(x_1,x_2)=4.2$ and EMD$(y_1,y_2)=2.8$.
Our algorithm separates the distributions effectively, with $\max D_\mathrm{KL} = 0.01$. 
In the middle panel (square), the simulated distributions again overlap fully; they have similar variances but different covariances, with EMD$(x_1,x_2)=1.3$ and EMD$(y_1,y_2)=1.1$.
The relabeled distributions appear rather accurate, although $\max D_\mathrm{KL}$ rises to $0.08$.
In the bottom panel (triangle), the simulated distributions overlap fully and have similar variance and covariance, with EMD$(x_1,x_2)=0.4$ and EMD$(y_1,y_2)=1.6$.
The relabeled distributions appear reasonably accurate, although $\max D_\mathrm{KL}$ rises to 0.3. 
In this case $\sim 40\%$ of the samples are mislabeled, but this happens mostly in overlapping regions where the two distributions are truly indistinguishable; as a result, the divergence is not affected too greatly.

Altogether, our algorithm is efficient in resolving the ambiguity between simple distributions, with moderate (and expected) degradation as overlaps increase, resulting in greater source confusion.

%%%%%%%%%%%%%%%%%%%%%%%%%%%%%%%%%%%%%
\subsection{Superimposed sinusoidal signals}
\label{sec:toy-sinusoid}
%%%%%%%%%%%%%%%%%%%%%%%%%%%%%%%%%%%%

\begin{figure*}
    \centering
    \includegraphics[width=\linewidth]{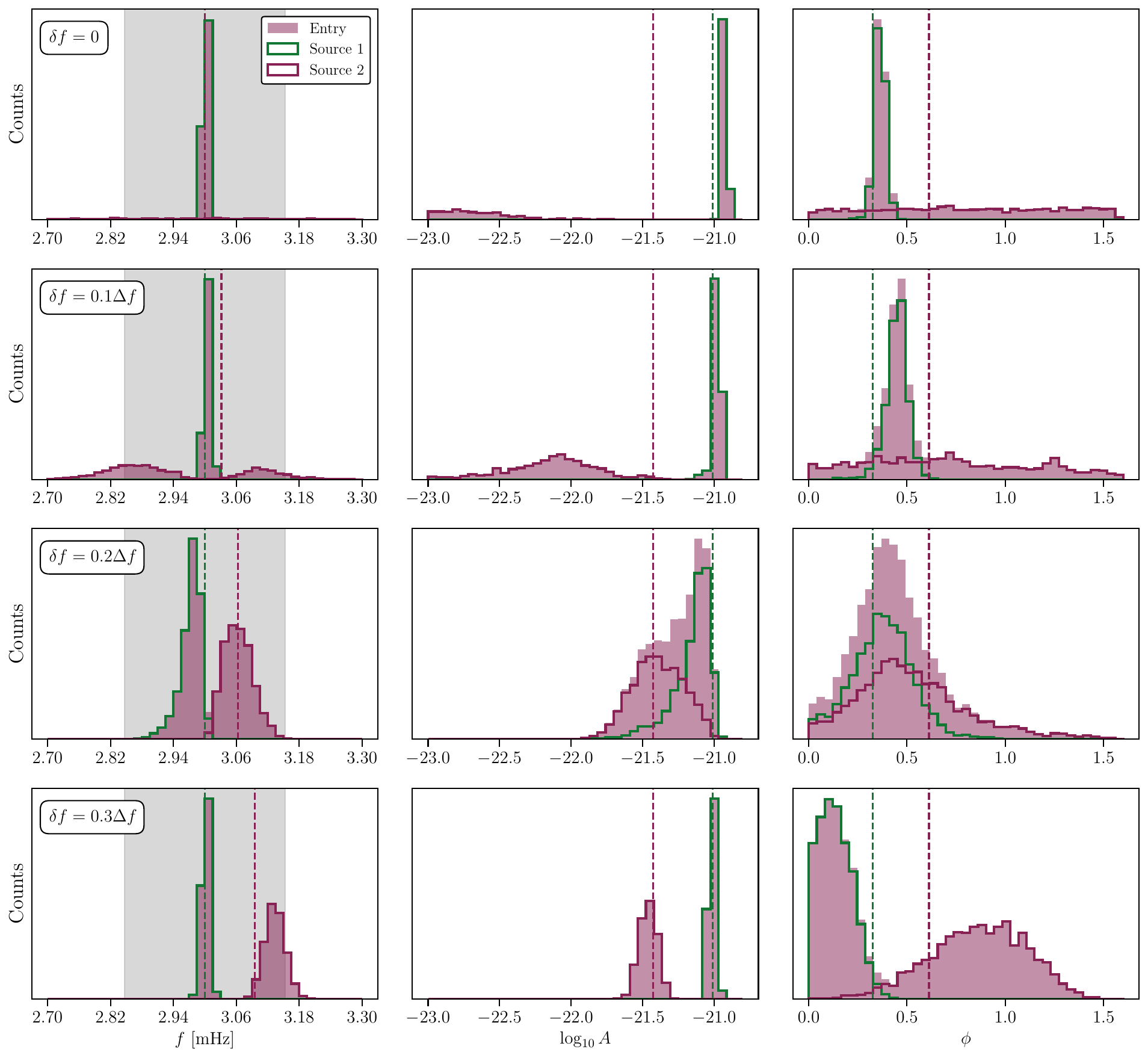}
    \caption{\textsc{Petra} relabeling of two superimposed sinusoids (second test of Sec.\ \ref{sec:toy-gaussian}).
    Each row corresponds to a different $\delta f$ between the sinusoids, and shows distributions of frequency (left), amplitude (middle), and phase (right).
    The gray shaded region covers a frequency bin.
    The filled magenta histogram displays all global-fit entries together, approximating the identical marginal distributions of the global fit.
    The magenta and green contours displays the two source catalog posteriors built by \textsc{Petra}. 
    Dashed vertical lines indicate the true (simulated) values of each parameter. 
    The two sources begin to be resolved at a frequency separation of 0.2 frequency bins.}
    \label{fig:frequency_sequence}
\end{figure*}

\begin{figure*}
    \centering
    \includegraphics[width=\linewidth]{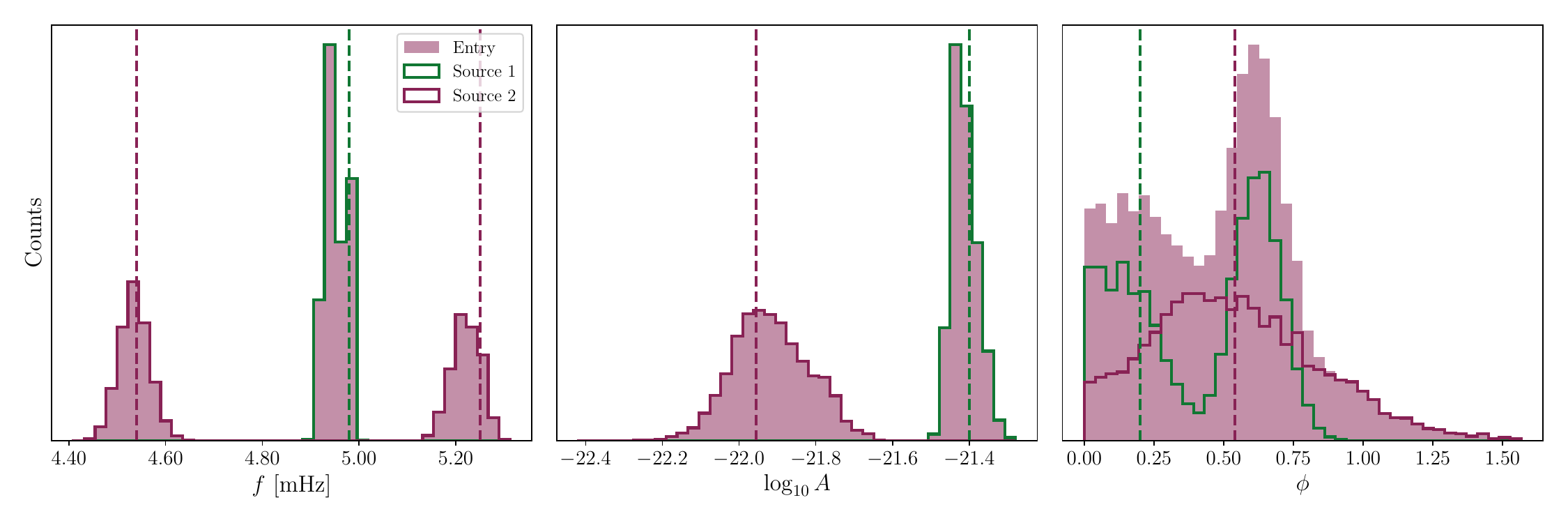}
    \caption{Same as Fig.~\ref{fig:frequency_sequence}, except that source 2 has a bimodal frequency posterior, emulated by simulating three signals (two with identical amplitude and phase but different frequencies) and recovering two. 
    \textsc{Petra} identifies each source correctly and recovers parameters consistent with their true values.}
    \label{fig:multimodal_sines}
\end{figure*}

Our second test is a simplified version of the LISA Galactic-binary global fit.
We analyze a time series $d(t_i = i \Delta T)$ that contains noise plus $N$ sinusoidal signals:
\begin{equation}
    \label{eq:sumsines}
    d(t_i) = n(t_i) + \sum_{\alpha=1}^N s_\alpha(t_i)\,,
\end{equation}
with $n(t_i) = n_i \sim {\cal{N}}(0, \sigma^2)$ with known $\sigma$, and $s_\alpha(t) = A_\alpha \sin(2 \pi f_\alpha t + \phi_\alpha)$.
Each signal depends on three parameters: an amplitude $A_\alpha$, a constant frequency $f_\alpha$, and a phase offset $\phi_\alpha$.
Equation~\eqref{eq:sumsines} implies that the likelihood is invariant with respect to parameter-dependent source-index permutations.
We simulate data with $\sigma = 5 \times 10^{-22}$, $\Delta T = 10$\,s, and timespan $T=10^{-4}$\,yr.
We adopt LISA-relevant priors $f \sim U(0.001 \, \mathrm{Hz}, 0.01 \, \mathrm{Hz})$, $\log_{10}A \sim U(-24, -20)$, and $\phi \sim U(0, 2 \pi)$. We sample the global-fit posterior using \textsc{impulse\_mcmc} \cite{johnson_aarondjohnsonimpulse_mcmc_2025}.
A rewrite of \textsc{PTMCMCSampler} \cite{Ellis:2017b}, \textsc{impulse\_mcmc} is an adaptive parallel-tempering Metropolis--Hastings Markov chain Monte Carlo sampler focusing on modularity and improved efficiency.
When estimating parameters for a variable number of sources, we sample from the transdimensional posterior using the product-space method \cite{Lodewyckx:2011a}.
We consider three problems under this general setup.

\paragraph*{Two sinusoids, with increasing frequency offsets.}
We begin by simulating and recovering two sources at nearby frequencies, investigating the impact of the frequency difference on source resolution.
The first source has $f_1= 0.003$ Hz, $\log_{10}A_1 =-21.008$, and $\phi_1=0.127$ with a signal-to-noise ratio\footnote{The single-source $\mathrm{SNR}$ is 
\begin{equation*}
    \mathrm{SNR}^2 = 4\int \frac{|\tilde{h}(f)|^2}{\sigma^2} \mathrm{d}f\,,
\end{equation*}
where $\tilde{h}(f)$ is the Fourier transform of the signal.} (SNR) of $27.6$.
The second source has $f_2 = f_1 + \delta f$ with $\delta f \in \{0, 0.1/T, 0.2/T, 0.3/T\}$, $\log_{10}A_2 =-21.427$, and $\phi_2=1.212$, with SNR $= 10.5$. 

The global-fit and \textsc{Petra} results are shown in Fig.~\ref{fig:frequency_sequence} for increasing $\delta f$ (top to bottom) when creating a catalog of 2 sources.
The filled purple histograms show the global-fit marginal posteriors for each parameter (they are statically identical for both signals).
The unfilled green and purple histograms show the posterior entries assigned by \textsc{Petra} to the two catalog sources.
The vertical dashed lines show the location of the true (simulated) parameters.

For $\delta f=0$ (first row), the two signals are fundamentally indistinguishable, since the sum of two sinusoids of the same frequency is a single sinusoid. 
The two-source catalog requested from \textsc{Petra} contains one source with the correct frequency, higher combined amplitude, and biased phase; and a second source with very low (undetectable) amplitude and unconstrained frequency and phase.
The situation is similar for $\delta f = 0.1/T$.
The two sources can be resolved when $\delta f = 0.2/T$ (20\% of a Fourier bin's width), and parameter recovery improves further for $\delta f = 0.3/T$.
\textsc{Petra}'s source resolution does not rely solely on frequency: experimenting with two sinusoidal signals with identical amplitude and phase, we find that they can be resolved only at $\delta f = 0.4/T$.

\begin{figure}
    \centering
    \includegraphics[width=\columnwidth]{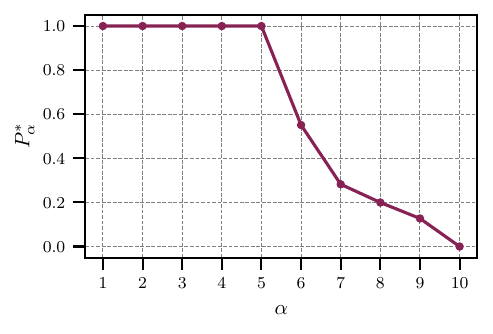}
    \caption{Probability of astrophysical origin for each of the 10 catalog sources, ordered by decreasing probability. The simulated data contain 10 sources, 5 of which are detectable, while the transdimensional global fit contains posterior samples with as little as 5 and as many as 9 sources.
    The \textsc{Petra} catalog reports $P^*_\alpha = 1$ for the 5 indisputable sources, lower values for the ``rogue'' global-fit entries, and  $P^*_\alpha = 0$ for a putative 10th source.}
    \label{fig:pastro_10_sinusoids}
\end{figure}

\begin{figure*}
    \centering
    \includegraphics[width=\linewidth]{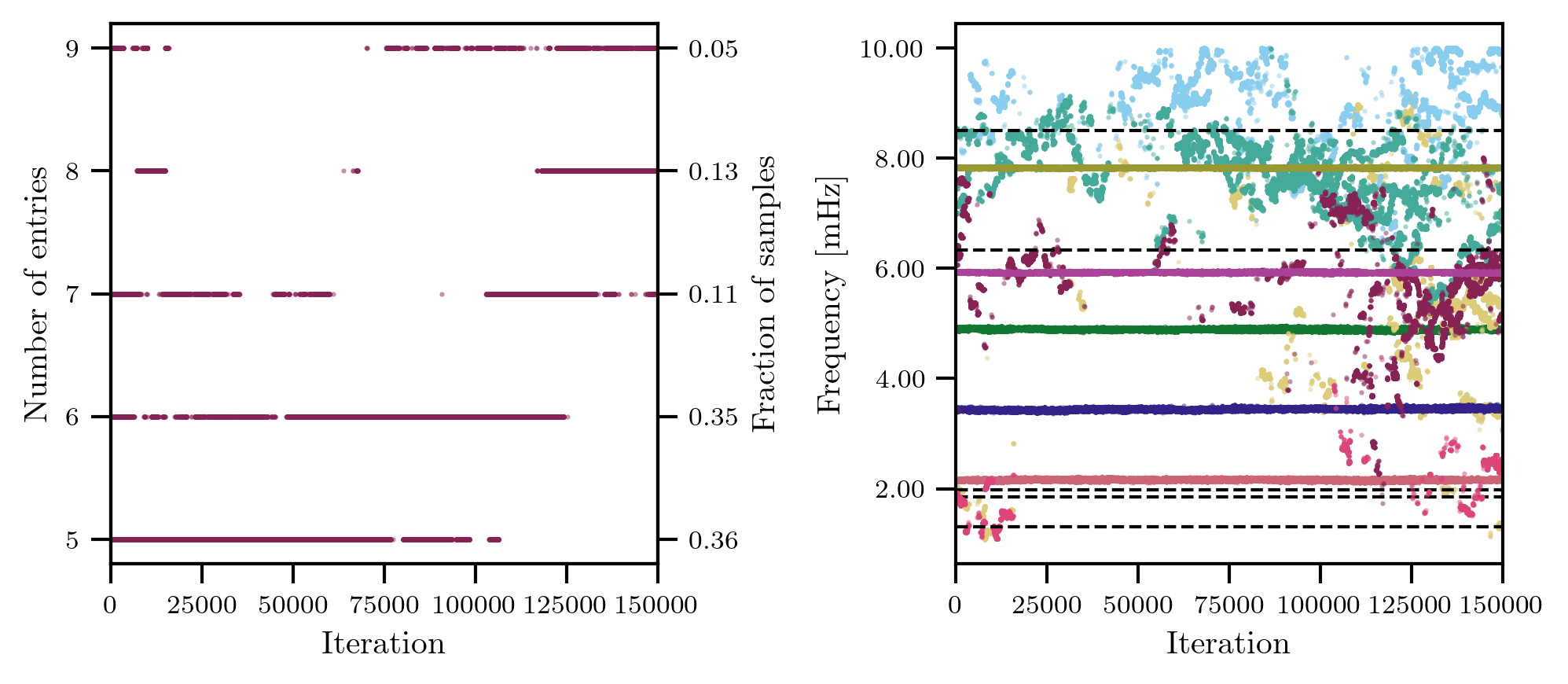}
    \caption{\textsc{Petra} relabeling of ten superimposed sinusoids (five detectable).
    \emph{Left panel}: number $N^i$ of global-fit entries across posterior samples, with the corresponding total fraction on the right vertical axis.  
    \emph{Right panel}: frequencies assigned to each catalog source. 
    Horizontal dashed lines show the true frequencies of the undetectable signals.
    The five detectable sources are clearly identified as such, while ``rogue'' global-fit samples are assigned to four catalog sources with $P^*_\alpha < 1$.
    The tenth catalog source is never used.}
    \label{fig:ten_sine_rj}
\end{figure*}

\paragraph*{Two sinusoids, with a multimodal posterior.}
We next consider the case where the posteriors are not only overlapping, but also multimodal, which tests the adequacy of the simple normal catalog ansatz of Eq.~\eqref{eq:gaussian}.
Bimodality is in fact encountered in global fits for LISA Galactic binaries~\cite{Littenberg:2020bxy}.
To emulate this situation with the sinusoidal signal setup, we simulate three signals (two of which have identical amplitude and phase but different frequencies) and recover only two.
Figure~\ref{fig:multimodal_sines} shows the global-fit marginal posteriors and the entries assigned by \textsc{Petra} to two catalog sources, with the same conventions as Fig.\ \ref{fig:frequency_sequence}.
The unimodal and bimodal sources are discriminated correctly, especially for frequency and amplitude---again thanks to the multiparameter operation of the algorithm.
Some confusion remains in the recovery of phase, which is likely related to \textsc{Petra} fitting the two effective frequencies of source 2 with a single broad normal ansatz.
A more flexible ansatz capable of bimodality would have improved source resolution further.

\paragraph*{Ten sinusoids, five detectable.}

Moving closer to the complexity of the LISA global fit, we simulate ten sources: five are detectable with SNR $>3$, while the rest have SNR $<1$.
Frequency and phases are drawn across the corresponding priors.
We sample the global fit using a transdimensional model with prior $N \sim U(0,10)$. 
The posterior chain reaches a maximum of $N=9$, and we construct a catalog of ten sources.
Figure~\ref{fig:pastro_10_sinusoids} shows the probability $P^*_{\alpha}$ that each source is real.
The five detectable sources are indisputable, with $P^*_{\alpha}=1$; they are included in every posterior chain sample.
The astrophysical probability decreases for the next four sources, indicating that they are present in only a fraction of the global fit.
Finally, the tenth source has $P^*_{\alpha}=0$ because no posterior sample contains ten sources, so \textsc{Petra} did not assign any entries to that catalog source.
In fact \textsc{Petra} returns sensible results even when asked to create inadvisably large catalogs.

The colored tracks in the right panel of Fig.~\ref{fig:ten_sine_rj} show the frequency of the posterior samples that are assigned to each of the ten catalog sources.
The five clear horizontal tracks correspond to the five detectable signals.
Beyond these, the global fit explores a number of additional signals, resulting in ``rogue'' traces that span the frequency prior, and do not lock onto the undetectable signals (the dashed lines in Fig.~\ref{fig:ten_sine_rj}).
The rogue entries are assigned to catalog sources somewhat randomly, although there is a degree of frequency clustering.

The left panel of Fig.~\ref{fig:ten_sine_rj} displays the number $N^i$ of entries active at each iteration of the sampler.
Five and six entries are preferred with roughly equal probabilities, followed by seven and eight, with nine entries active in only 5\% of the samples. These shares are in general agreement with the $P^*_\alpha$ of Fig.~\ref{fig:pastro_10_sinusoids}.
Note that $p_\mathrm{gf}(N = 9|d) = 0.05$ is not directly comparable with $P^*_9 \sim 0.2$, since different catalog sources capture different rogue entries along the global-fit chains.

%%%%%%%%%%%%%%%%%%%%%%%%%%%%%%%%%%%%%
\section{Application to mock LISA data}
\label{sec:globalfit}
%%%%%%%%%%%%%%%%%%%%%%%%%%%%%%%%%%%%

In this section we consider the problem of building catalogs from LISA global fits of Galactic binaries, experimenting with simulated data over a small frequency band.
We obtain global fits with the \textsc{UCBMCMC} sampler \cite{ldasoft}, which separates the full LISA frequency band into smaller frequency segments that are analyzed individually.
\textsc{UCBMCMC} uses a reversible-jump, parallel-tempering MCMC on each segment, sampling the joint posterior for the number of sources and their parameters~\cite{Littenberg:2020bxy, Littenberg:2023xpl}.
We consider a single segment with ten simulated signals, analyzing 0.25 years of data with no noise.
This short duration reduces the \textsc{UCBMCMC} computational cost, but it makes separating overlapping sources more challenging, because it reduces the range of the sky-location--dependent Doppler shifts induced by the LISA orbital motion.

Each Galactic-binary signal depends on the eight parameters $\theta = (f , \dot{f}, A, \lambda, \cos \vartheta , \cos \iota, \psi, \phi)$ with $f$ the frequency, $A$ the amplitude, $\lambda$ the longitude, $\vartheta$ the colatitude, $\iota$ the inclination, $\psi$ the polarization, and $\phi$ the phase offset \cite{Littenberg:2020bxy}. 
With stationary Gaussian noise, the optimal single-source SNR may be computed for Galactic binaries using the noise-weighted inner product $\rho^2 = \langle \tilde{h}(f) \mid \tilde{h}(f)\rangle$, per Eq.~(14) of Ref.~\cite{Littenberg:2020bxy}.
The larger number of parameters compared to the toy problem of Sec.~\ref{sec:toy-sinusoid} should help \textsc{Petra} resolve overlapping signals more efficiently.
\begin{figure}
    \centering
    \includegraphics[width=\columnwidth]{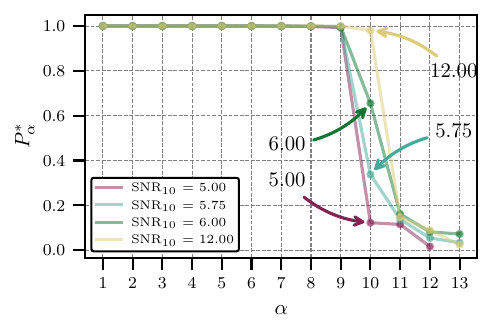}
    \caption{Probability of astrophysical origin $P^*_\alpha$ for each catalog source in a simulated LISA dataset containing ten Galactic binary signals with well separated frequencies. Nine signals are expected to be detectable with $\mathrm{SNR} > 15$; they are identified confidently in our catalog, with $P^*_{\alpha}=1$.
    The probability of astrophysical origin of the tenth source increases nonlinearly with its SNR.}
    \label{fig:ucb_pastro}
\end{figure}

We draw the true (simulated) source parameters from the LISA Data Challenge \textsc{Sangria} dataset \cite{LeJeune:2022a}, but modify source frequencies to be evenly separated, and adjust the amplitudes. 
Nine of the ten signals have SNR $> 15$, while we vary the SNR of the 10th signal ($\mathrm{SNR}_{10}$) between 5 and 12.
We analyze the data with \textsc{UCBMCMC} and obtain a posterior chain that includes samples with as many as 15 entries.
We then run \textsc{Petra} to construct a catalog with the maximum number of entries.
Figure~\ref{fig:ucb_pastro} shows the probability of astrophysical origin for each source, for different values of $\mathrm{SNR}_{10}$. 
The probability is $P^*_{\alpha}=1$ for the first nine sources, which are clearly detectable. 
The probability $P^*_{10}$ that the 10th source is real increases nonlinearly with $\mathrm{SNR}_{10}$, from 0.12 at $\mathrm{SNR}_{10}=5$, to 0.66 at $\mathrm{SNR}_{10}=6$, to 0.98 at $\mathrm{SNR}_{10}=12$.  
Sources 11 to 13 remain at $P^*_{\alpha}<0.2$.
\begin{figure}
    \centering
    \includegraphics[width=\columnwidth]{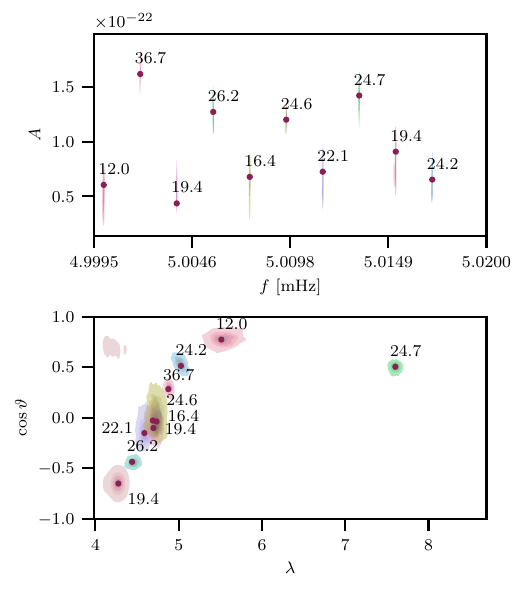}
    \caption{
    Frequency + amplitude (top) and longitude + colatitude (bottom) catalog posteriors for the 10-binary simulated LISA dataset.
    Dots mark the true (simulated) parameters;
    shaded colored regions denote posterior contours for each source as a different color, and are annotated with SNRs; magenta points indicate true values.}
    \label{fig:ucb_params}
\end{figure}

Figure~\ref{fig:ucb_params} shows catalog parameter posteriors for the ten confidently recovered catalog sources when $\mathrm{SNR}_{10}=12$.
The top panel shows frequency and amplitude, while the bottom panel shows sky location.
\textsc{Petra} resolves a large number of label switches in the original global-fit chain. 
% These are both inherent to how the global fit posterior was sampled in the first place and the entry shuffling we perform at initialization.
All distributions are centered around the true parameter values; they are well separated for frequency, but have significant overlaps in sky position. The pink source with $\mathrm{SNR}=19.4$ displays a bimodal posterior, which is handled correctly by \textsc{Petra} even if the auxiliary catalog distribution is Gaussian. 
\begin{figure}
    \centering
    \includegraphics[width=\columnwidth]{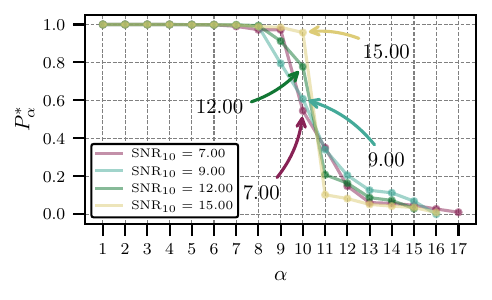}
    \caption{Same as Fig.~\ref{fig:ucb_pastro}, but for datasets in which sources 9 and 10 have frequencies within two Fourier bins. We fix $\mathrm{SNR}_9 = 15$ and vary $\mathrm{SNR}_{10}$ from 7 to 15. 
    The frequency overlap confuses the identification of sources 9 and 10, creating an additional source 11 for the lower values of $\mathrm{SNR}_{10}$.}
\label{fig:ucb_pastro_close_signals}
\end{figure}

We repeat the same analysis after modifying the dataset so that two signals have overlapping frequency posteriors. 
We move the leftmost (lowest-frequency) signal in Fig.~\ref{fig:ucb_params} to within two frequency bins of the next signal. These two signals are assigned by \textsc{Petra} to catalog sources 10 and 9, respectively, so we will use these indices to refer to them. We adjust $\textrm{SNR}_{9}$ from 36.7 to 15, and vary $\textrm{SNR}_{10}$ from 7 to 15.

Figure~\ref{fig:ucb_pastro_close_signals} shows the catalog probabilities of astrophysical origin for different values of $\mathrm{SNR}_{10}$.
The eight well separated, loud signals are resolved easily.
When $\mathrm{SNR}_{10}=15$, sources 9 and 10 are recovered confidently, and no further signals emerge with $P^\ast_\alpha$ greater than 0.2.
For lower values of $\mathrm{SNR}_{10}$, source confusion hampers identification; $P^\ast_{10}$ and (less so) $P^\ast_{9}$ decrease accordingly, while a spurious source 11 emerges, reaching $P^\ast_{11} \simeq 0.4$ for $\mathrm{SNR}_{10}=7$.
\begin{figure*}
    \centering
    \includegraphics[width=\textwidth]{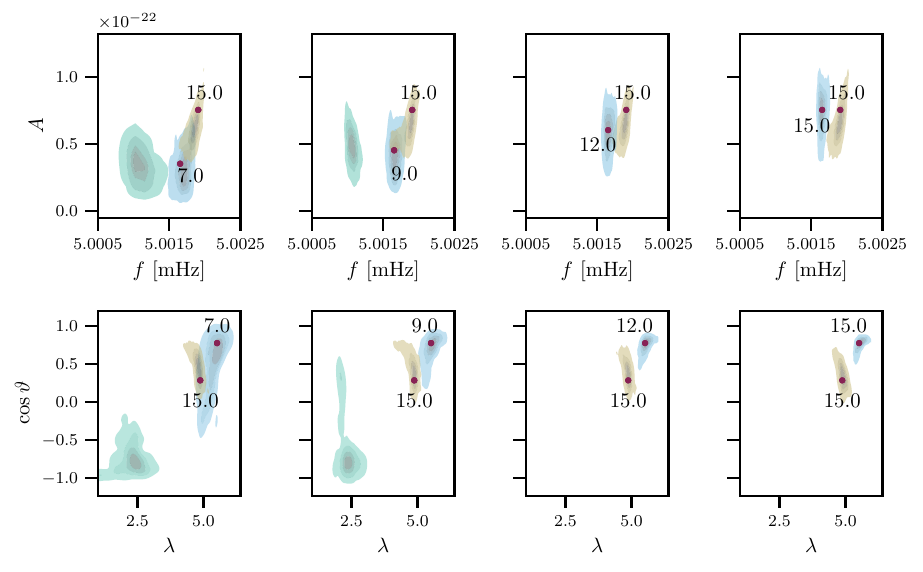}
    \caption{Amplitude + frequency and longitude + colatitude catalog posteriors for sources 9 (yellow), 10 (blue), and 11 (green, when $P^*_{11} > 0.2$),
    for datasets in which sources 9 and 10 have frequencies within two Fourier bins.
    Results for $\mathrm{SNR}_{10} = 7$, 9, 12, and 15 are shown in the four columns.}
    \label{fig:ucb_params_close_signals}
\end{figure*}

These trends are explored in Fig.~\ref{fig:ucb_params_close_signals}, which shows catalog posteriors for sources 9--11.
When $\mathrm{SNR}_{10} \ge 12$ (the two rightmost panels), sources 9 and 10 are separated confidently and have unimodal posteriors.
For lower $\mathrm{SNR}_{10}$ (the two leftmost panels), the catalog posterior of source 9 (yellow) remains stable; the posterior of source 10 (blue) widens, but remains unimodal and consistent with the true parameters; and remarkably, the global fit populates an additional entry with similar parameters, which \textsc{Petra} assigns to source 11 (green), with $P^*_{11} \simeq 0.3$--$0.4$.

\begin{figure}
    \centering
    \includegraphics[width=\columnwidth]{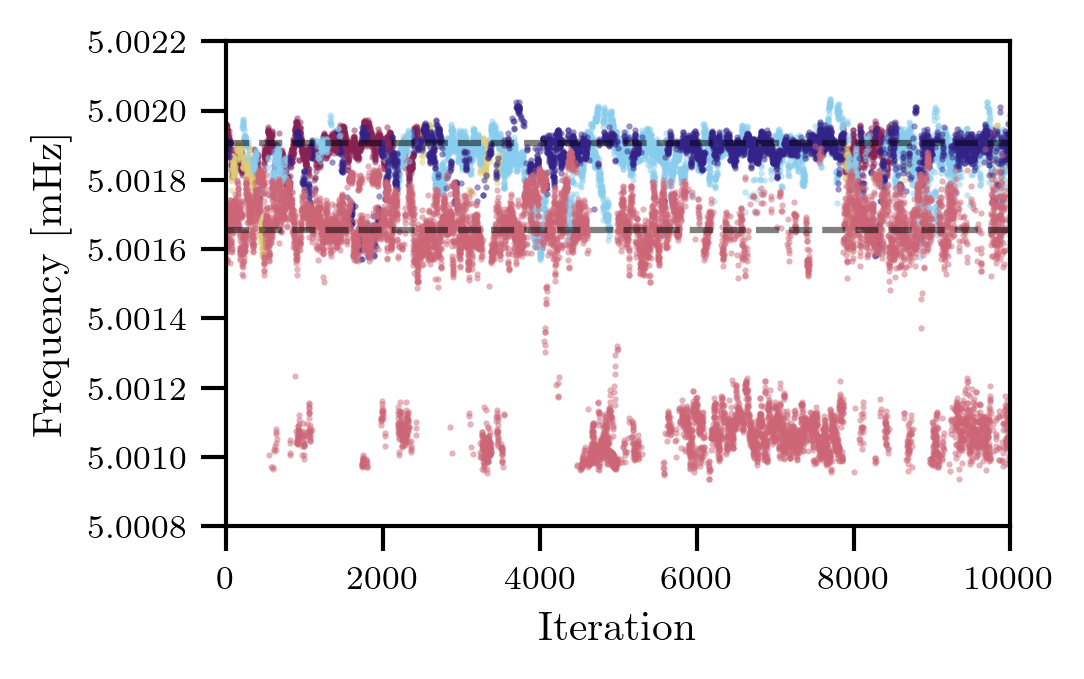}
    \caption{Global-fit frequency traces for a dataset in which signals 9 and 10 have frequencies within two Fourier bins, with $\mathrm{SNR}_{10} = 9$ (cf.\ the second column of Fig.\ \ref{fig:ucb_params_close_signals}).
    Traces are thinned by a factor 20, and the frequency range is restricted to catalog sources 9--11.
    The variety of colors shows that the same catalog source occurs across different global-fit entries. Dashes show the true (simulated) frequencies of signals 9 and 10. The cluster around 5.001 mHz occurs across $\sim$ 40\% of the samples and is assigned to catalog source 11.}
    \label{fig:overlapping_chain_example}
\end{figure}

Figure~\ref{fig:overlapping_chain_example} explores the emergence of source 11 by plotting global-fit frequency chains in the neighborhood of sources 9 and 10, for the $\mathrm{SNR}_{10}=9$ dataset.
The colors correspond to sampler indices; as expected, the same catalog source is represented across different entries.
Dashed lines show the true frequency values of the two simulated sources in this frequency range.
Posterior samples cluster broadly around three frequencies: two of them correspond to the true frequencies, while the third may result from interference.

Almost none (0.5\%) of the global-fit entries assigned to source 11 come from samples where all three frequencies are represented, which forces \textsc{Petra} to assign the additional source.
Most of the source-11 assignments come at the expense of source 10, so \textsc{Petra} is instead choosing to split a bimodal posterior into two catalog sources. 
Equation~\eqref{eq:reward} implies a tradeoff between a greater $P^*_\alpha$ for a single source with a broader posterior and lower $P^*_\alpha$ for two sources with narrower posteriors.
In this case it is possible that \textsc{Petra} is constrained by its simple Gaussian ansatz for the approximate catalog posteriors, which cannot conform to a bimodal posterior.

Note that our algorithm is deterministic \emph{given a global-fit chain}; different initial arrangements of global-fit entries may lead to different local maxima of the catalog reward function.
% The two sources of uncertainty in this process are (a) the initial shuffling of the global fit entries per sample, and (b) whether the algorithm converges to a single global optimum or oscillates between different relabelings while still maximizing the assignment probability.
To study this aspect, we repeat the analysis of Fig.~\ref{fig:ucb_pastro_close_signals} by reshuffling the global fit entries randomly.
The results presented above correspond to the reshuffling that yields the lowest cost, Eq.~\eqref{eq:log_joint_prob}.
Nevertheless, the total reward remains consistent to four decimal places, and the astrophysical probabilities of very likely and very unlikely sources are also highly stable. Source assignments are very stable for the former, very unstable for the latter (which is neither surprising nor harmful).
The largest variation in $P^*_\alpha$ appears for medium-confidence sources: for $\mathrm{SNR}_{10}=9$, $P^\ast_{10}$ varies between 0.5 and 0.9, corresponding to a stronger and weaker source 11, respectively.
% This variation results from the catalog preferring two $P^\ast\sim0.5$ sources in higher-reward catalogs over a single $P^\ast\sim0.9$ source in lower-reward catalogs.
% To study the latter, for the $\mathrm{SNR}_{f}=9$ case, we inspect the progression of how entries are assigned to different sources. 
% High-$P^\ast$ sources are stable, while the ``rogue" samples keep getting swapped between low-$P^\ast$ sources, even close to catalog convergence.
% Since these sources are unlikely to be real, we still conclude that the confident sources are robustly extracted.

Overall, \textsc{Petra} successfully creates a catalog from a transdimensional global fit over simulated LISA data, even when signals have overlapping frequency support.

%%%%%%%%%%%%%%%%%%%%%%%%%%%%%%%%%%%%
\section{Discussion}
\label{sec:conclusions}
%%%%%%%%%%%%%%%%%%%%%%%%%%%%%%%%%%%

In this article we have presented a novel method to construct a source catalog from a global-fit solution.
Our work is motivated by the LISA global-fit problem for Galactic-binary signals~\cite{Littenberg:2020bxy}, but it generalizes to any inference problem where an unknown number of indistinguishable signals (i.e., signals that have identical functional dependence on the same set of parameters) are combined additively in the data. 
This creates a label-switching ambiguity whereby the global fit posterior is invariant to permutations of the parameters among sources (for a fixed total number of sources). 
A traditional astronomical catalog is instead a set of independent single-source posteriors and their probabilities of being real, which are not invariant to label switching.

The LISA global-fit problem is complicated by this \emph{label-switching ambiguity}; by the \emph{transdimensional nature} of inference, given that the number of indistinguishable sources is not known in advance; and by the \emph{confusion} that can occur between true signals with similar parameters, especially for weaker signals at the threshold of detection.
All three aspects come to the fore in the creation of a catalog.

The \emph{label-switching ambiguity} is resolved by a circular (but not tautological) process.
We seek the source-parameter-dependent \emph{relabeling} (which maps each set of label-switching-equivalent global-fit parameters to a single ordered subset of catalog sources) such that the relabeled global-fit posterior is statistically closest to the product of the single-source catalog distributions.
The latter are defined as the marginal posteriors of the relabeled global fit.
We implement this process by approximating the catalog probabilities with parametric auxiliary distributions, and then iteratively optimizing the relabeling and the auxiliaries to minimize the divergence between the relabeled global fit and the auxiliaries.

The \emph{variable number of global-fit sources} is handled by defining the catalog as associating to each source both its posterior distribution \emph{and} its probability of being real.
Global-fit entries are assigned to catalog sources on the basis of their parameter posteriors and also of their astrophysical probability.
Assignments that place many (few) samples in sources with a low (high) probability of being real are penalized naturally in our method. 

\emph{Source confusion} means that the distillation of the full global fit to a catalog will necessarily lose information, as the latter cannot represent any parameter correlations among sources.
Nevertheless, catalogs even if imperfect are desirable in astronomical and astrophysical investigations that target individual sources rather than populations: for instance, searches for electromagnetic counterparts, joint analyses of gravitational-wave signals and light curves, and more.

Our method is described in Sec.\ \ref{sec:methods}, and it is implemented in the open-source software package \textsc{petra\_catalogs}~\cite{johnson_petra_2025}. In Secs.\ \ref{sec:toy} and \ref{sec:globalfit} we demonstrate its operation with two toy models and with a simulated LISA global fit.
\textsc{Petra} displays robust source-resolving performance even when single-source posteriors overlap significantly, or are multimodal in one of the parameters.
Among \textsc{Petra}'s benefits:
\begin{itemize}
    \item It works in postprocessing using the output (typically, a posterior chain) of any global-fit sampler, leaving the sampling procedure intact.
    \item It handles transdimensional posteriors without discarding samples.
    \item It returns a catalog of sources that contains both parameter posteriors and astrophysical probability.
    \item For simplicity, it relies on simple Gaussian auxiliary distributions, but it can be adapted easily to more sophisticated alternatives. 
    \item Most important, it makes use of the full posterior over all signal parameters, thus exploiting all available information to distinguish sources.
\end{itemize}

%%%%%%%%%%%%%%%%%%%%%%%%%%%%%%%%%%%%%%%%
\paragraph*{Comparison to past work.}
\citet{Umstatter:2005su} explored a toy model involving a large number of overlapping sinusoids in an early effort to demonstrate LISA data analysis.
To separate signals, they examined windows around the peaks of the frequency posterior to evaluate the presence of multiple sinusoids.
If those were found, they performed hierarchical Ward clustering~\cite{Ward01031963} across all three sinusoid parameters under the condition that each entry in a global-fit sample had to be assigned to a different cluster.
This form of clustering begins with one cluster for every entry and sample, and then combines them iteratively to minimize within-cluster variance (based on Euclidian parameter distances).
The final result is a single cluster per source, which defines catalog entries.
In effect, this algorithm augments inference with a source-identification criterion based on clustering in parameter space, without taking into account the structure of signal space or the presence of instrument noise.

\citet{Littenberg:2020bxy} (also \cite{Littenberg:2023xpl,Lackeos:2023eub}) performed label reassignment using a heuristic clustering procedure that relies on waveform similarity in signal space rather than parameter posteriors.
Waveform similarity is assessed with the normalized noise-weighted inner product (the \emph{match}~\cite{Owen:1995tm,LIGOScientific:2019hgc}).
A catalog source
is defined as the collection of global-fit entries whose waveforms exceed a chosen match threshold, by default 0.5~\cite{Littenberg:2020bxy}.
While \textsc{Petra} tends to group ``rogue'' samples together in sources with wide distributions (see, e.g., Fig.~\ref{fig:ten_sine_rj}), Ref.~\cite{Littenberg:2020bxy} introduces a new source for each entry that does not exceed the match threshold with other sources.
Like \textsc{Petra}, this method is also applied in postprocessing and it implicitly uses all source parameters. However, because it associates sources by comparing their noiseless waveforms, this method does not directly account for the noise realization in the actual data.

Last, \citet{Buscicchio:2019rir} introduced a generic method to resolve the label-switching ambiguity during sampling, adopting a volume-preserving, invertible map from the full parameter hypercube (which contains $N!$ copies of the posterior) to a single hypertriangle.
This map preserves the prior.
Compared to \textsc{Petra}, however, this method relies on a single parameter to create the map.

All four methods are expected to yield similar results for easily detectable, nonoverlapping sources: a comparison for sources at the edge of detectability is left for future work.

\paragraph*{Caveats and further work.}
The \textsc{Petra} framework can be extended as more complex global-fit scenarios are considered.
First, the auxiliary distributions are multivariate Gaussians.
Though this ansatz can still separate sources with non-Gaussian posteriors, it is undoubtedly suboptimal.
More flexible distributions would be straightforward to implement.
An intriguing choice would be normalizing flows~\cite{Kobyzev:2021a}, which could be included in the catalog as synthetic representations of the marginal posteriors, allowing both posterior draws and posterior evaluation for arbitrary parameters.

Second, LISA data will be analyzed by multiple teams that will create different global fits.
The ultimate astrophysical catalog therefore needs to be created by combining multiple global fits directly, or by combining their catalogs.
We are interested in extending our approach to this problem.

Third, LISA will see many types of sources.
Extending the framework presented here to cover each ``spoke'' of the global-fit ``wheel'' should be straightforward due to the lack of label-switching symmetry between different source types: the global fit decides which sources fall under each source type, and our algorithm relabels the posterior separately for each subcatalog.
However, some source types may not pose any label-switching problem because all signals are clearly distinct.

Last, LISA data analysis will be performed continuously as more data are obtained and downlinked throughout the duration of the mission.
Thus, global fits and catalogs will evolve with time. Catalogs in particular will need to account for sources that appear, disappear, or simply become better constrained as more data are analyzed.
We expect that all these extensions will be possible within the \textsc{Petra} framework, and plan to pursue them in future investigations.

%%%%%%%%%%%%%%%%%%%%%%%%%%%%%%%%%%%%%%
\begin{acknowledgments}
%%%%%%%%%%%%%%%%%%%%%%%%%%%%%%%%%%%%%%

We thank Tyson Littenberg for assistance with \textsc{UCBMCMC}, which was used to simulate and analyze LISA data. We thank Pat Meyers, Marco Crisostomi, Jonah Kanner, and Curt Cutler for helpful discussions.
This project was kickstarted at the second LISA sprint, which was hosted by Caltech and supported by the Jet Propulsion Laboratory Astronomy and Physics Directorate.
We acknowledge support from the Caltech and Jet Propulsion Laboratory President and Director's Fund (ADJ, KC, and MV), from the Sloan Foundation (ADJ and KC), and from the NASA LISA Study Office (MV).
JR acknowledges support from the Sherman Fairchild Foundation.
KAG acknowledges support from an NSF CAREER grant \#2146016.
\\
Software: \textsc{SciPy}~\cite{Virtanen:2019joe}, \textsc{NumPy}~\cite{harris2020array}, \textsc{JAX}~\cite{jax2018github}, \textsc{Pandas}~\cite{the_pandas_development_team_2024_13819579, mckinney-proc-scipy-2010}, \textsc{matplotlib}~\cite{Hunter:2007}, and \textsc{Seaborn}~\cite{Waskom2021}.
\end{acknowledgments}

\bibliography{apssamp}% Produces the bibliography via BibTeX.

\end{document}